\documentclass{mn2e}
\usepackage{natbib}
\usepackage{graphicx}
\usepackage{pslatex}

\begin{document}


\title[Spectra from a leaky box]{Radio synchrotron spectra for a leaky box approximation}

\author[C.R. Kaiser]{C. R. Kaiser\thanks{crk@soton.ac.uk}\\ 
School of Physics \& Astronomy, University of Southampton, Southampton SO17 1BJ
}

\maketitle

\begin{abstract}
The synchrotron emission observed from many astrophysical objects is commonly modelled to arise from relativistic electrons in an emission region occupied by a spatially homogeneous, but tangled magnetic field. However, we know that magnetic fields embedded in ionised gases tend to form flux ropes interspersed with regions of much lower magnetic field strengths. Here we develop a full description of the evolution of the energy distribution of relativistic electrons in a plasma divided into two distinct regions with different strengths of the magnetic field. Electrons are able to continuously leak from the low-field region into the high-field region. The model becomes fully analytic for physically reasonable assumptions. We show that such a leaky box model produces two distinct breaks in the electron energy distributions which give rise to three breaks in the resulting synchrotron spectrum. The spectral slopes in between the breaks are in general not constant and thus allow for significant curvature of the spectrum. These spectra are consistent with spatially resolved observations of the radio spectra of the lobes of radio galaxies. The exact form of the spectra depends on the adopted diffusion rate. The leaky box model significantly extends the time over which synchrotron emission can be detected at a given frequency compared to the usually assumed case of homogeneous magnetic fields. The spectral ages inferred for the electron population from standard techniques for the leaky box model are considerably younger than their real age.
\end{abstract}


\section{Introduction}

Radio synchrotron radiation arises from relativistic electrons moving in magnetic fields. The very nature of the synchrotron process enables us to infer the energy distribution of the electrons from the observed radio spectrum \citep[e.g.][]{nk62,ap70}. In many astrophysical objects this energy distribution is a power-law of the electron energy which arises from the first order Fermi acceleration at shock fronts in gas flows \citep[e.g.][]{ab78}. Over time, energy losses of the electrons due to the emission of synchrotron radiation as well as to the inverse Compton scattering of photons, mainly from the Cosmic Microwave Background (CMB) radiation, modify the shape of the energy distribution by introducing a high-energy cut-off \citep[e.g.][]{rl79}. This high-energy cut-off gives rise to an exponential drop of the synchrotron spectrum at high frequencies. The position in frequency space of this drop is used to infer the `spectral age' of the electron population, i.e. the time the electrons have spent in the magnetic field since they were accelerated to relativistic speeds. 

The standard spectral ageing technique assumes that the magnetic field is homogeneous throughout the emitting volume and that its strength is constant for the spectral age of the electron population \citep[e.g.][]{sb77,www80,al87,pa87,lpr92}. Adiabatic expansion of the emitting volume does not change the shape of the electron energy distribution, but moves the high-energy cut-off to lower energies as well as decreases the frequency of the associated spectral break. Consequently, the inferred spectral ages are modified, if adiabatic expansion is taken into account \citep{ba94}. 

Other effects that make spectral age estimates problematic are in-situ re-acceleration of the relativistic electrons and inhomogeneous magnetic fields. Indications that these effects may play an important role were already reported by \citet{pa87}. The observation of radio spectra of spatially resolved lobes of radio galaxies with significantly more `curvature'  than expected from the standard spectral ageing model emphasized these concerns \citep{kra93,rka94}. While re-acceleration clearly remains a possible explanation for the observations, we concentrate in this paper on the effects of an inhomogeneous magnetic field. 

For Cygnus A \citet{kr94} found that the significantly curved radio spectra from any location within the radio lobes, including the jets and hotspots, could be traced back to a single spectral shape by shifting the observed spectra in $\log \left({\rm flux} \right)$--$\log \left( {\rm frequency} \right)$ space. This result led \citet{br00} to develop a model for powerful radio galaxies in which a gradient in the magnetic field strength along the lobes is responsible for  `illuminating' different parts of a uniform electron population throughout the lobes. It is not clear how the specific gradient in the magnetic field required by the model can be maintained and the required conditions for the very fast diffusion of the relativistic electrons through the lobes may not apply \citep{ck00b}. Thus models with a magnetic field inhomogeneous on scales of the radio lobes may not explain the observations. 

On the other hand, it is well known that the strength of the magnetic field in the solar plasma is far from homogeneous on small scales \citep[e.g.][]{cz87} and the observations of filamentary structure in the lobes of radio galaxies \citep[e.g.][]{cpdl91} may confirm the existence of similar flux ropes in these structures. \citet{wg90} and \citet{sw90} develop a model with two separate regions of different magnetic field strengths which make up the synchrotron emitting plasma. The electron populations within the two regions also remain separate. They show that, when observed with a telescope unable to resolve the two regions, such a configuration can significantly alter the radio spectrum compared to the case of a homogeneous magnetic field and hence complicates estimates of the spectral age. 

In the case where the strength of the magnetic field in the emitting volume follows a Gaussian distribution, the radio spectrum does not develop an exponential break in the GHz range \citep{pt93,pt94}. Radiative energy losses of the electrons lead to a steepening of the radio spectrum, but since the loss history of individual electrons is very different, no clear break develops at low frequencies. The overall spectrum shows significant curvature\footnote{The predictions of the models presented in \citet{pt93} and \citet{pt94} for the curvature of the spectrum never appeared in print, but can be found at http://www.rfcgr.mrc.ac.uk/$\sim$ptribble/papers/preprints/color}, consistent with the observations of \citet{kra93} and \citet{rka94}. 

The idea of a synchrotron emitting volume with two regions of different magnetic field strengths was revisited by \citet{emw97} and \citet{ata00}. In the first work the relativistic electrons are allowed to either move diffusively through the plasma or leak from one distinct region into the other. Both approaches lead to radio spectra with more than a single, exponential break. Thus they predict significant deviations of the estimated spectral ages from the real ages of the electron population. However, the mathematical complexity of these models makes them difficult to study their predictions in detail or to apply them to observational data. 

In this paper we develop a simplified version of the leaky box model which is based on original ideas of the transfer of cosmic rays from the Galactic disc to the Galactic halo \citep[e.g.][]{nk62,gs64}. The main difference between our model and the work of \citet{emw97} and \citet{ata00} is that the rate at which electrons leak from the region with a low magnetic field strength into the region with a high magnetic field strength is independent of the electron energy. While this is a great simplification of the physics, it allows us to formulate an analytical solution for the evolution of the electron energy distribution. We are then able to investigate the predictions of this simple leaky box model in some detail and find that it explains many of the observed properties of radio synchrotron spectra.

Throughout the paper we assume that pitch angle scattering for the relativistic electrons is efficient and that a uniform pitch angle distribution is therefore maintained at all times. This behaviour implies that our model is of JP-type \citep{jp73}, rather than KP-type \citep{nk62,ap70}.

In Section \ref{develop} we derive expressions for the energy distribution of the electrons in both the low-field and high-field regions without making assumptions about the exact evolution of the volumes of or magnetic field strengths inside these regions as a function of time. In Section \ref{simple} we make further simplifications and introduce an explicit time-dependence to all relevant quantities which then allows the construction of a fully analytical solution. We then use this analytical solution in Section \ref{spectra} where we discuss the implications of the model for radio synchrotron spectra and for the standard spectral ageing estimates. We summarise our results in Section \ref{conc}.

\section{Development of the model}
\label{develop}

In our model the plasma containing the relativistic particles and the magnetic field is split into two separate, but spatial connected regions. The first region, region 1 or the `low-field' region, of volume $V_1(t)$ contains only a weak magnetic field with an energy density $u_{\rm B1}(t)$. Correspondingly, the radiative energy losses of relativistic electrons in this region will be small. The strength of the magnetic field in the second region, region 2 or the 'high-field' region, of volume $V_2(t)$ is greater than that in region 1 so that $u_{\rm B2}(t) > u_{\rm B1} (t)$. The radiative energy losses of the electrons in this region will be larger than those in region 1, but the synchrotron emissivity of region 2 will also be greater. 

The volumes of regions 1 and 2 may depend on time. Thus the energy densities of the magnetic fields and the energy distribution of the relativistic electrons, even in the absence of radiative losses, will also be functions of time. For simplicity we assume that the ratio $V_1(t) / V_2(t)$ remains constant. It is straightforward to adapt our results to more complicated cases where $V_1$ and $V_2$ have different temporal behaviour.

In our model the relativistic electrons contained in region 1 are allowed to leak into region 2. We do not consider leakage in the reverse direction. The main motivation for this assumption is the mathematical tractability of the problem. With leakage from region 2 back into region 1, we cannot formulate an analytic solution to the relevant equations. Leakage in the reverse direction is also less interesting for practical reasons. Given that we expect most of the observable radiation to be produced in the high-field region and that the electrons will have a relatively short lifetime in this region, leakage from region 2 into region 1 will mainly influence the low energy end of the electron energy distribution in region 1. The relevant electrons will not contribute to the emission at frequencies we normally observe. Finally, the gyro-radius of a given electron will be smaller in region 2 compared to region 1. It will therefore be more difficult for electrons to diffuse from region 2 to region 1 compared to diffusion in the opposite direction.

\subsection{Electron distribution in the low-field region}
\label{region1}

The electron number density in region 1 is governed by the kinetic equation of form
\begin{equation}
\frac{\partial n_1 \left( \gamma, t \right)}{\partial t} = \frac{\partial}{\partial \gamma} \left[ F\left( \gamma , t \right) n_1 \left( \gamma, t \right) \right] - \left[ 3 g\left( t \right) + h_1 \left( t \right) \right] n_1 \left( \gamma, t \right),
\label{kinetic}
\end{equation}
where
\begin{equation}
g\left( t \right) = \frac{1}{3 V_1 \left( t \right)} \, \frac{\partial V_1 \left( t \right)}{\partial t}
\label{volex}
\end{equation}
describes the dilution of the electrons by the expanding volume. We express the energy dependence of the electron density evolution in terms of the relativistic Lorentz factor, $\gamma$. The function $F\left( \gamma , t \right)$ contains all processes which affect the energy of individual electrons. 

The leakage of electrons from region 1 into region 2 is described by the sink term $h_1 n_1$. Note that the diffusion rate, $h_1$, does not depend on $\gamma$. In other words, of all electrons with a certain Lorentz factor $\gamma$ at time $t$, a fixed fraction diffuses during a time interval ${\rm d} t$. The size of this fraction does not depend on $\gamma$. In general $h_1$ will depend on the energy of the diffusing electrons. However, we will show in the following that this restriction is necessary for an analytical solution of equation (\ref{kinetic}). Usually we will assume that the energy distribution of the relativistic electrons in region 1 follows, at least initially, a power-law in $\gamma$ with a negative exponent. Thus there are many more electrons at low energies compared to high-energy electrons. Consequently, more low-energy electrons will diffuse into region 2. These low-energy electrons accumulate in region 2 and will eventually contribute significantly to the overall pressure inside this region while their contribution to the pressure in region 1 diminishes. At this point we would expect that further diffusion of low-energy electrons is suppressed, thereby introducing an energy dependence in $h_1$. At this point our solution derived below becomes unphysical. However, again only the low-energy end of the electron distribution is affected which does not influence the synchrotron emissivity at common observing frequencies. So for practical purposes we can ignore this effect. 

We consider two forms of energy loss for individual relativistic electrons. The adiabatic expansion of $V_1$ not only dilutes the electrons, but electrons also lose energy due to the work done during the expansion. The adiabatic loss term is given by $g\left( t \right) \gamma$ \citep[e.g.][]{ml94}. Radiative energy losses due to synchrotron radiation are proportional to $\gamma^2$, if the magnetic field is tangled on scales much smaller than the dimensions of the volume occupied by the plasma. The same proportionality holds for energy losses due to inverse Compton scattering of CMB photons. We can therefore write for these combined loss processes $f_1 \left( t \right) \gamma ^2$. All energy losses are then given by
\begin{equation}
F\left(\gamma, t \right) = f_1\left( t \right) \gamma^2 + g\left( t \right) \gamma.
\end{equation}

By noting that
\begin{equation}
\frac{\partial n_1 \left( \gamma, t\right)}{\partial t} = \frac{{\rm d}n_1 \left( \gamma , t \right)}{{\rm d} t} - \frac{\partial n_1 \left( \gamma, t \right)}{\partial \gamma} \frac{\partial \gamma}{\partial t},
\end{equation}
we can now separate equation (\ref{kinetic}) into two new equations. The evolution of the Lorentz factor of individual electrons is described by
\begin{equation}
\frac{\partial \gamma}{\partial t} = - f_1 \left( t \right) \gamma ^2 - g\left( t \right) \gamma,
\label{difgam}
\end{equation}
which has the well-known solution \citep[e.g.][]{nk62}
\begin{equation}
\gamma = \frac{\gamma _0 G\left( t, 0 \right)}{\gamma _0 \int _0^t G\left( t', 0 \right) f_1 \left( t' \right)\, {\rm d}t' +1},
\label{gamevol1}
\end{equation}
with
\begin{equation}
G \left( t, t_{\rm i} \right) = \exp \left[ - \int_{t_{\rm i}}^t g \left( t' \right) \, {\rm d} t' \right].
\label{Gdef}
\end{equation}
$\gamma _0$ is the Lorentz factor at $t=0$ of electrons with Lorentz factor $\gamma$ at time $t$. Note that the high-energy cut-off, $\gamma _{0,{\rm max}}$, for the initial distribution injected into region 1 at time $t=0$, translates into a cut-off at 
\begin{equation}
\gamma _{\rm 1,max} \left( t \right)= \frac{\gamma _{0,{\rm max}} G\left( t, 0 \right)}{\gamma _{0,{\rm max}} \int _0^t G\left( t', 0 \right) f_1 \left( t' \right)\, {\rm d}t' +1}
\label{cutoff}
\end{equation}
at all later times $t>0$. This high-energy cut-off is present even if $\gamma _{0,{\rm max}} \rightarrow \infty$.

The second equation resulting from the separation of equation (\ref{kinetic}) describes the evolution of the electron density as a whole as
\begin{equation}
\frac{{\rm d} n_1 \left( \gamma , t \right)}{{\rm d} t} = \left[ 2 \gamma f_1 \left( t \right) - 2 g \left( t \right) - h_1 \left( t \right) \right] n_1 \left( \gamma , t \right).
\end{equation}
Because we restrict $h_1$ to be independent of $\gamma$, we can solve this differential equation by separation of variables to give 
\begin{eqnarray}
n_1\left( \gamma, t \right) & = & n_{1,0} \exp \left\{ \int _0 ^t  \left[ 2 \gamma f_1 \left( t' \right) -2 g \left( t' \right) -h_1 \left( t' \right) \right] \, {\rm d} t' \right\} \nonumber\\
& & H \left( \gamma_{\rm 1,max} - \gamma \right),
\end{eqnarray}
where $n_{1,0}$ is the initial electron energy distribution injected into region 1 at $t=0$. The Heaviside function $H$ arises from the high-energy cut-off of the electron distribution, see equation (\ref{cutoff}). From equation (\ref{difgam}) we note that
\begin{equation}
\gamma f_1 \left( t \right) = -g \left( t \right) - \frac{\partial \ln \gamma}{\partial t},
\end{equation}
and upon substituting for $\gamma f_1$, separating the integral in the exponential and using equation (\ref{gamevol1}) we find
\begin{eqnarray}
n_1 \left( \gamma, t \right) & = & n_{1,0} \left( \frac{\gamma _0}{\gamma} \right)^2 G^4 \left( t, 0 \right) \exp \left[ - \int _0^t h_1 \left( t' \right) \, {\rm d} t' \right] \nonumber\\
& & H \left( \gamma_{\rm 1,max} - \gamma \right).
\label{dist1}
\end{eqnarray}

\subsection{Electron distribution in the high-field region}

We will now derive an expression for the electron energy distribution in region 2, $n_2 \left( \gamma , t \right)$. Consider electrons leaking from region 1 into region 2 during a short time interval $\Delta \tau$ at a time $\tau$ with $0\le \tau \le t$. The energy distribution of these electrons in region 1 is given by $\Delta n_1 \left( \gamma _{\tau}, \tau \right) = n_1 \left( \gamma _{\tau} , \tau \right) h_1 \left( \tau \right) \Delta \tau$. The Lorentz factor of the electrons at the time of leaking into region 2 is $\gamma _{\tau}$. Particle conservation requires $V_1 \left( \tau \right) h_1 \left( \tau \right) = V_2 \left( \tau \right) h_2 \left( \tau \right)$ and also $\Delta n_1 \left( \gamma _{\tau}, \tau \right) V_1 \left(\tau \right) = \Delta n_2 \left( \gamma _{\tau}, \tau \right) V_2 \left( \tau \right)$. Thus the energy distribution in region 2 of the electrons crossing from region 1 into region 2 at $t=\tau$ is given by
\begin{equation}
\Delta n_2 \left( \gamma _{\tau}, \tau \right) = n_1 \left( \gamma _{\tau}, \tau \right) h_2 \left( \tau \right) \Delta \tau.
\label{Dn2}
\end{equation}

After arrival in region 2, the energy distribution of the electrons is governed by a kinetic equation similar to equation (\ref{kinetic}), but without a sink term,
\begin{eqnarray}
\frac{\partial}{\partial t} \Delta n_2 \left( \gamma , t \right) & = & \frac{\partial}{\partial \gamma} \left\{ \left[ f_2 \left( t \right) \gamma ^2 + g\left( t \right) \gamma \right] \Delta n_2 \left( \gamma , t \right) \right\} \nonumber\\
& & - 3 g \left( t \right) \Delta n_2 \left( \gamma , t \right).
\end{eqnarray}
We use the same separation technique as in Section \ref{region1}. The evolution of the Lorentz factor of individual electrons in region 2 is then given by
\begin{equation}
\gamma = \frac{\gamma _{\tau} G \left( t, \tau \right)}{\gamma _{\tau} \int _{\tau}^t G \left( t', \tau \right) f_2 \left( t' \right) \, {\rm d} t' +1}.
\label{gamevol2}
\end{equation}
The high-energy cut-off for the electrons contributing to $\Delta n_2 \left( \gamma , t \right)$ can be found by substituting $\gamma _{\tau,{\rm max}}$ for $\gamma _{\tau}$ in this expression. Clearly, $\gamma _{\tau, {\rm max}}$ is the high-energy cut-off of the electrons leaking from region 1 into region 2 at time $\tau$ and thus can be determined from equation (\ref{cutoff}) by replacing $t$ with $\tau$.

The energy distribution of the electrons is governed by
\begin{equation}
\frac{{\rm d}}{{\rm d} t} \Delta n_2 \left( \gamma , t \right) = \left[ 2 \gamma f_2 \left( t \right) - 2 g \left( t \right) \right] \Delta n_2 \left( \gamma , t \right),
\end{equation}
which has the solution
\begin{equation}
\Delta n_2 \left( \gamma , t \right) = \Delta n_2 \left( \gamma _{\tau}, \tau \right) \left( \frac{\gamma _{\tau}}{\gamma} \right)^2 G^4 \left( t, \tau \right),
\end{equation}
where we have used equation (\ref{gamevol2}).

Using equation (\ref{Dn2}), we can replace $\Delta n_2 \left( \gamma _{\tau}, \tau \right)$ with an expression containing the short time interval $\Delta \tau$. The complete energy distribution of all electrons leaking from region 1 into region 2 can then be calculated by replacing $\Delta \tau$ with the differential ${\rm d} \tau$ and integrating,
\begin{equation}
n_2 \left( \gamma , t \right) = \int _{\tau _{\rm min}}^t \left( \frac{\gamma _{\tau}}{\gamma} \right)^2 G^4 \left( t, \tau \right) n_1 \left( \gamma _{\tau} ,\tau \right) h_2 \left( \tau \right) \, {\rm d} \tau.
\label{dist2}
\end{equation}
The lower integration limit is governed by the evolving high-energy cut-off of the relativistic electrons in region 2, 
\begin{equation}
\gamma _{\rm 2,max} \left( t, \tau \right)= \frac{\gamma _{{\rm 1,max}} \left( \tau \right)  G \left( t, \tau \right)}{\gamma _{{\rm 1,max}} \left( \tau \right) \int _{\tau} ^t G \left( t', \tau \right) f_2 \left( t' \right) \, {\rm d} t' +1}.
\end{equation}
Substituting for $\gamma _{1, {\rm max}} \left( \tau \right)$ from equation (\ref{cutoff}) and rearranging then gives
\begin{equation}
\gamma _{\rm 2,max} \left( t, \tau \right)= \frac{\gamma _{0,{\rm max}} G \left( t, 0 \right)}{\gamma _{0, {\rm max}} \mathcal{G} + 1},
\label{imp}
\end{equation}
where
\begin{equation}
\mathcal{G} = G \left( \tau, 0 \right) \int _{\tau}^t G \left( t', \tau \right) f_2 \left( t' \right) \, {\rm d} t' + \int _0^{\tau} G\left( t', 0 \right) f_1 \left( t' \right) \, {\rm d} t' .
\end{equation}
For a given Lorentz factor $\gamma$ we expect, at least at early times $t$, that $\gamma \le \gamma_{\rm 2,max}$ and in this case $\tau _{\rm min} = 0$ in equation (\ref{dist2}). At later times radiative energy losses will have moved the high-energy cut-off, $\gamma _{\rm 2,max}$, to Lorentz factors below $\gamma$. In this case, equation (\ref{imp}) is an implicit equation for the integration limit $\tau_{\rm min}$. This integration limit also removes the need for the Heaviside function in the definition of $n_1 \left( \gamma _{\tau}, \tau \right)$ in equation (\ref{dist1}).

In addition to the electrons leaking from region 1 into region 2, some electrons may have been injected into region 2 directly at time $t=0$. The energy distribution of these electrons will have the functional form of equation (\ref{dist1}) with $n_{1,0}$ and $h_1$ replaced by $n_{2,0}$ and $0$ respectively. Also, the evolution of the Lorentz factor of individual electrons is given by equation (\ref{gamevol2}) with $\tau =0$ and $\gamma _{\tau} = \gamma _0$.

\section{The simple leaky box}
\label{simple}

In this section we will simplify the general expressions for the electron distribution functions derived above by suitable choices for a number of the parameters involved in the model. The aim is to build an analytical model which is simple to use, but allows us to study the generic properties of the synchrotron emission from a leaky box. 

\subsection{Initial energy distributions and diffusion rates}

The exact functional form of the injected electron energy distributions, $n_{0,1}$ and $n_{0,2}$, depends on the mechanism responsible for the electron acceleration. Here, we will not consider distribution functions with a contribution from an initial injection of electrons into region 2. The synchrotron emission of such electrons is short-lived compared to that of electrons leaking from region 1 into region 2. At times where $\gamma _{\rm 2,max}  \left( t, 0 \right)< \gamma$ for electrons with Lorentz factor  $\gamma$ giving rise to synchrotron emission at observable frequencies, there is no contribution left from electrons initially injected into region 2. 

For electron acceleration at shock fronts, the most likely scenario for the lobes of radio galaxies of radio-loud quasars, we expect $n_{0,1} = n_0  \gamma _0^{-p}$ with $2 \le p < 3$ and a constant $n_0$ \citep[e.g.][]{ab78,hd88}. By choosing $p=2$ we can significantly simplify the expressions for $n_1$ and $n_2$ and for the purpose of illustrating the basic properties of the leaky box model, we restrict our further discussion to this case. There are no analytic solutions of the various integrals in the expressions for $n_1$ and $n_2$ for $p \ne 2$. However, the only effect of choosing $p>2$ is a general steepening of the spectral slopes discussed below. 

From equation (\ref{dist1}) we then find
\begin{equation}
n_1 \left( \gamma , t \right) = n_0 \gamma ^{-2} G^4 \left( t, 0 \right) \exp \left[ - \int _0^t h_1 \left(t' \right) \, {\rm d} t' \right] H \left( \gamma_{\rm 1,max} - \gamma \right).
\end{equation}
The electron distribution in region 2 from equation (\ref{dist2}) simplifies to
\begin{equation}
n_2 \left( \gamma , t \right) = n_0 \gamma ^{-2} G^4 \left( t, 0 \right) \int_{\tau_{\rm min}}^t h_2 \left( \tau \right) \exp \left[ - \int_0^{\tau} h_1 \left( t' \right)\,{\rm d} t' \right] \, {\rm d} \tau.
\end{equation}

The simplest diffusion rate we can adopt is $h_1 = {\rm constant}$. We have already made the assumption that the time dependence of $V_1$ is the same as that of $V_2$. Thus we also have $h_2 = {\rm constant}$ from particle conservation. This choice then results in
\begin{equation}
n_1 \left( \gamma , t \right) = n_0 \gamma ^{-2} G^4 \left( t, 0 \right) \exp \left( -h_1 t \right) H \left( \gamma_{\rm 1,max} - \gamma \right),
\label{simn1}
\end{equation}
and
\begin{equation}
n_2 \left( \gamma , t \right) = n_0 \gamma ^{-2} G^4 \left( t , 0 \right) \frac{h_2}{h_1} \left[ \exp \left( - h_1 \tau_{\rm min} \right) - \exp \left(-h_1 t \right) \right].
\label{simn2}
\end{equation}

\subsection{Explicit time dependence}

The time dependence of all relevant quantities in the model is fixed by our choice for $V_1 \left( t \right)$ and $V_2 \left( t \right)$ as functions of time. We will concentrate on $V_1 \left( t \right)$ bearing in mind our earlier assumption that the behaviour of $V_2 \left( t \right)$ as a function of $t$ is the same. 

The evolution of $V_1$ is governed by the fluid dynamics the plasma of the leaky box is subject to. In many practical applications of the model we will have $V_1 \propto t^a$ with $a \ge 0$ and so we set
\begin{equation}
V_1 \left( t \right) = V_{1,0} \left( 1 + \frac{t}{t_0} \right)^a,
\end{equation}
where $V_{1,0}$ is the volume at $t=0$ and $t_0$ is a `scale height' determining the speed of expansion of the volume. Substitution into equation (\ref{volex}) yields
\begin{equation}
g \left( t \right) = \frac{a}{3} \frac{1}{t+t_0},
\end{equation}
while equation (\ref{Gdef}) gives
\begin{equation}
G \left( t , t_{\rm i} \right) = \left( \frac{t + t_0}{t_{\rm i} + t_0} \right)^{-a/3}.
\end{equation}

The assumption of a magnetic field tangled on small spatial scales allows us to treat the field simply as a magnetic pressure with equivalent energy density $u$ and adiabatic index $\Gamma = 4/3$. For our two regions we then have $u_1 \left( t \right)\propto u_2 \left( t \right) \propto V_1 ^{-4/3} \left( t \right)$. We set $u_{\rm j} \left( t \right) = u_{{\rm j},0} \left( 1 + t/t_0 \right)^{-4a /3}$, where the index ${\rm j}$ can be either 1 or 2. $u_{1,0}$ and $u_{2,0}$ are constants and by definition $u_{1,0} < u_{2,0}$. Thus the radiative energy losses are described by
\begin{equation}
f_{\rm j} \left( t \right) = \frac{4}{3} \frac{\sigma _{\rm T}}{m_{\rm e} c} \left[ u_{\rm j} \left( t \right) +u_{\rm CMB} \right],
\end{equation}
where $\sigma _T$ is the Thomson cross-section, $m_{\rm e}$ is the electron rest mass, $c$ the speed of light and $u_{\rm CMB}$ the energy density of the CMB. A useful result is the solution of the integral
\begin{eqnarray}
\int_{t_{\rm i}}^t G \left( t', t_{\rm i} \right) f_{\rm j} \left( t' \right) \, {\rm d} t' & = & \frac{4}{a} \frac{\sigma _{\rm T}}{m_{\rm e} c}  \left( t_{\rm i} + t_0 \right) \nonumber \\
& & \left\{ \frac{u_{\rm j} \left( t_{\rm i} \right)}{b} \left[ 1-G^b \left(t,t_{\rm i} \right) \right] + \right. \nonumber \\
& & \left. \frac{u_{\rm CMB}}{d} \left[ 1- G^d \left( t, t_{\rm i} \right)  \right] \right\}, 
\end{eqnarray}
with 
\begin{eqnarray}
b & =& 5-3/a \nonumber\\
d & = & 1-3/a. 
\end{eqnarray}

\section{Radio spectra of the simple leaky box}
\label{spectra}

With the results from Section \ref{simple} we can now study the properties of the radio spectra arising from the simple leaky box model. We compare these with the spectral properties of radio galaxies. We therefore adopt model parameters appropriate for these objects. Note that the behaviour of the model may be quite different for a different choice of the model parameters. Here we only want to illustrate the general behaviour for the situation in radio galaxies.

The model for the dynamical evolution of the lobes of radio galaxies of type FRII \citep{fr74} by \citet{ka96b} predicts that volume elements of the lobe containing relativistic electrons injected at the same time evolve with $V \propto t^a$ with \citep{kda97a}
\begin{equation}
a=\frac{4+\beta}{\Gamma _{\rm c} \left( 5 + \beta \right)}.
\end{equation}
The model assumes a power-law distribution for the density of the gas the lobe is expanding into and $-\beta$ is the exponent of this distribution. In the following we use $\beta=1.5$. The adiabatic index of the lobe material, $\Gamma _{\rm c}$, is set to $4/3$, since we are only interested in the magnetic field and the relativistic electrons, both of which have a relativistic equation of state.

The expansion time of the lobes of radio galaxies is of order $10^7$\,years and so we set $t_0$ equal to that time. The strength of the magnetic field in the lobes of Cygnus A is estimated as $\sim 100$\,$\mu$G \citep{cpdl91} and so we set $B_2 \left( t=0 \right) = 100$\,$\mu$G for the strong-field region and $B_1 \left( t=0 \right) = B_2 \left( t=0 \right) /10$ in the low-field region. The initial energy densities of the magnetic fields in the two regions is given by $u_{\rm j,0} = B_j^2 \left(t=0 \right) / \left( 8 \pi \right)$. We neglect any initial injection of relativistic electrons into region 2 and so the initial pressure in  region 2 is $p_0= u_2 /3$. For pressure equilibrium between the two regions we require $p = \left( u_1 + u_{\rm e} \right) / 3$, from which we obtain $u_{\rm e}$, the initial energy density of the relativistic electrons in region 1. We assumed earlier a power-law energy distribution for the electrons injected into region 1 with an exponent $p=2$. Thus we have \citep{kda97a}
\begin{equation}
n_0 = \frac{u_{\rm e}}{m_{\rm e} c^2} \left[ \ln \left( \gamma _{0,\rm max} \right) + \frac{1}{\gamma_{0,\rm max}} -1 \right].
\end{equation}
We set $\gamma _{0,\rm max} =10^6$, but note that the exact value of this initial high-energy cut-off is not important. 

Finally, we choose the diffusion rate $h_1 = 0.2 / t_0$ and set $h_2 = h_1$. The latter statement implies that regions 1 and 2 have the same volume. The energy density of the CMB is $u_{\rm CMB} = 4.3 \times 10^{-13}$\,ergs\,cm$^{-3}$, equivalent to a negligible redshift.

These settings fix the density of relativistic electrons in both regions. The synchrotron emissivity, i.e. the luminosity per unit volume,  of each region is then given by
\begin{equation}
\epsilon _{\rm j, \nu} \left( t \right)= \int_1^{\gamma _{\rm j,max}} n_{\rm j} \left( \gamma , t \right) j_{\nu} \left( \gamma , t \right) \, {\rm d} \gamma,
\end{equation}
where $j_{\nu} \left( \gamma , t \right)$ is the emissivity of a single electron with Lorentz factor $\gamma$. In our calculations below we use the analytical expression for $j_{\nu}$ presented in \citet{ggs88}.

The total monochromatic radio luminosity of regions 1 and 2 combined is given by $P_{\nu} = \left( V_1 + V_2 \right) \epsilon _{{\rm tot}, \nu} = V_1 \epsilon _{1,\nu} +V_2 \epsilon_{2,\nu}$, where $\epsilon _{{\rm tot}, \nu}$ is the emissivity that would be inferred from observing $P_{\nu}$ under the assumption of a homogeneous emission region. It is interesting to note that 
\begin{equation}
\frac{V_1 + V_2}{V_1} \epsilon _{{\rm tot}, \nu} = \epsilon _{1,\nu} + \frac{V_2}{V_1} \epsilon_{2,\nu} = \epsilon _{1,\nu} + \frac{h_1}{h_2} \epsilon _{2,\nu}.
\end{equation}
From equations (\ref{simn1}) and ({\ref{simn2}) we then see that only the normalisation, but not the shape of the spectrum of the radio emission from the simple leaky box model depends on the ratio $V_2/V_1 = h_1/h_2$. In other words, the spectral shape depends only on the diffusion rate between the two regions, but is independent of their respective volume filling factors.

\subsection{Electron energy distributions and spectra}
\label{spec}

The radiative energy losses of the relativistic electrons in region 1 cause the high-energy cut-off of the electron energy distribution, $\gamma_{1,\rm max}$, to evolve according to equation (\ref{cutoff}). Since the radiative losses of electrons in region 2 are greater than in region 1 and because electrons constantly leak from region 1 into region 2, $\gamma_{1,\rm max}$ is also the high-energy cut-off of the electron energy distribution in region 2. However, $n_2$ also contains a break at $\gamma _{2,\rm max} \left(t,0 \right)$, defined by equation (\ref{imp}). For $\gamma \le \gamma _{2,\rm max} \left(t,0 \right)$, radiative losses have not yet moved the high-energy cut-off of electrons, which leaked into region 2 at a time $\tau =0$, below their current Lorentz factor. Thus the shape of the energy distribution is still that of the energy distribution initially injected into region 1. Also, note that $\tau_{\rm min} =0$ in this case. For Lorentz factors exceeding $\gamma _{2,\rm max} \left(t,0 \right)$, $\tau_{\rm min} > 0$ and electrons which leaked into region 2 at times $\tau < \tau_{\rm min}$ do not contribute anymore to the energy distribution. Due to this effect the shape of the energy distribution will be steeper between $\gamma_{2,\rm max} \left( t, 0 \right)$ and $\gamma _{1,\rm max}$ than below $\gamma_{2,\rm max} \left( t, 0 \right)$.

\begin{figure}
\centerline{
\includegraphics[width=8.45cm]{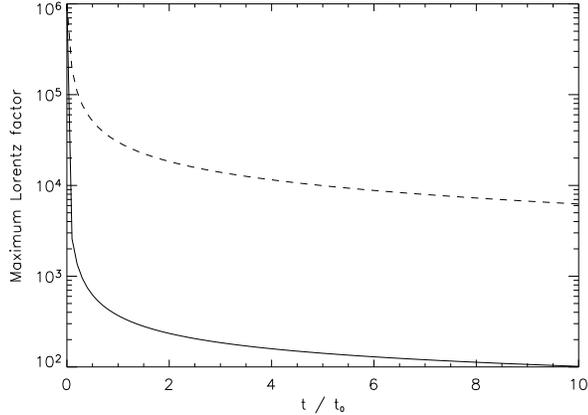}}
\caption{Evolution of the maximum Lorentz factor in the two regions. Solid line: $\gamma _{\rm 2,max} \left( t , 0 \right)$ in region 2 (high-field region). Dashed line: $\gamma _{\rm 1, max}$ in region 1 (low-field region).}
\label{gammax}
\end{figure}

Figure \ref{gammax} shows the evolution of the high-energy cut-off, $\gamma _{\rm 1, max}$, and of the break, $\gamma _{\rm 2, max} \left( t, 0 \right)$. The early evolution is very fast due to the strong energy dependence of the radiative losses. Later on the maximum Lorentz factors decrease only slowly. However, this does not imply that the resulting radio spectrum remains constant, because the strength of the magnetic field and the density of the relativistic electrons continue to decrease.

\begin{figure}
\centerline{
\includegraphics[width=8.45cm]{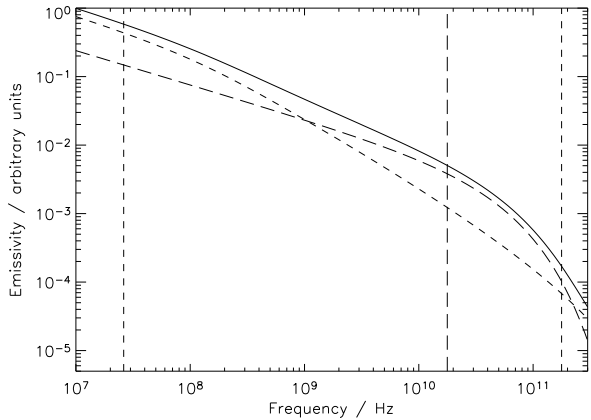}}
\caption{Radio spectrum of the simple leaky box at a time $t=0.5 t_0$. Long-dashed line: Contribution from region 1 (low-field region). Short-dashed line: Contribution from region 2 (high-field region). Solid line: Total spectrum of both regions added together. The vertical lines show the locations of the expected spectral breaks. Long-dashed, vertical line: Spectral break ($\nu_1$) in region 1 due to $\gamma _{\rm 1, max}$. Short-dashed, vertical lines: Spectral breaks in region 2 due to $\gamma _{\rm 2, max} \left( t, 0 \right)$ at low frequencies ($\nu_3$) and due to $\gamma _{\rm 1, max}$ at high frequencies ($\nu_2$).}
\label{sp05}
\end{figure}

As discussed above, the high-energy cut-off, $\gamma _{\rm 1, max}$, gives rise to two breaks in the radio spectrum of the leaky box, one in the contribution from region 1 (hereafter referred to as $\nu _1$) and the second in the contribution from region 2 (hereafter $\nu_2$). This is illustrated in Figure \ref{sp05}, which shows the radio spectrum at an early time, $t=0.5 t_0 = 5\times 10^6$\,years. At this point in time, about 10\% of all electrons initially injected into region 1 have leaked into region 2. Due to severe radiative losses in region 2, the break frequency associated with $\gamma _{\rm 2,max} \left( t, 0 \right)$ (hereafter $\nu_3$) has already moved to very low frequencies. We can approximately predict the location of spectral breaks by using the fact that the synchrotron emission of relativistic electrons with Lorentz factor $\gamma$ moving in a magnetic field of strength $B$ peaks around $\gamma^2 \nu_{\rm g} /3$, where the non-relativistic gyro-frequency is given by $\nu_{\rm g} = e B / \left( 2 \pi m_{\rm e} c \right)$ \citep[e.g.][]{rl79}.

The gyro-frequency of electrons in region 2 exceeds that in region 1 by an order of magnitude. Thus, the overall spectrum  at low frequencies is dominated by the contribution from region 2 where large numbers of low-energy electrons radiate at higher frequencies than their counterparts with the same Lorentz factor in region 1. Beyond the break frequency $\nu _3$ radiative losses in region 2 have decimated the number of electrons, the spectrum of region 2 steepens and the flat-spectrum contribution of region 1 dominates the overall emission. Note that due to the high frequency tail of the single electron emissivity, $j_{\nu}$, the spectra extend well beyond the break frequencies. This effect allows the contribution from region 2 to dominate the overall spectrum at frequencies higher than $\nu_2$. 

\begin{figure}
\centerline{
\includegraphics[width=8.45cm]{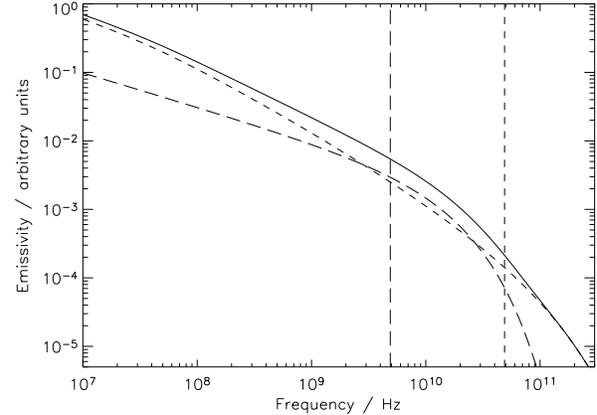}}
\caption{Same as Figure \ref{sp05}, but for time $t=t_0$. Note that the break at $\nu_3$ has now moved to frequencies below the lower limit shown here.}
\label{sp1}
\end{figure}

Figure \ref{sp1} shows the radio spectrum at $t=t_0$. Region 2 now contains 18\% of all initially injected electrons. This fraction is enough to allow the emission from region 2 to exceed that from region 1 at almost all frequencies. At frequencies between $\nu_3$ (not visible in the Figure) and $\nu_1$ the overall spectrum is steeper than the spectrum of region 1 alone due to the radiative losses in region 2. However, at frequencies above $\nu_1$ the overall spectrum is flatter than that of region 1. Also, the spectrum extends to higher frequencies, beyond the cut-off in the spectrum of region 1. 

\begin{figure}
\centerline{
\includegraphics[width=8.45cm]{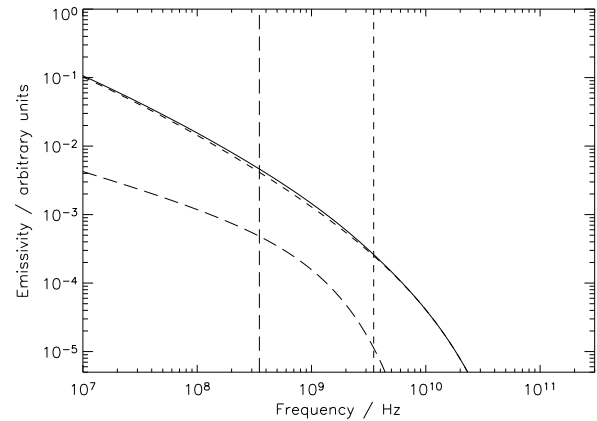}}
\caption{Same as Figure \ref{sp05}, but for time $t=4 t_0$.}
\label{sp4}
\end{figure}

Finally, Figure \ref{sp4} shows the radio spectrum of the simple leaky box at a time $t=4t_0$, when 55\% of all electrons initially in region 1 have leaked to region 2. The overall spectrum is now essentially identical to the spectrum emitted by the electrons in region 2. Although this spectrum is still steeper at frequencies below $\nu_1$ than that of region 1, it is clear that at frequencies above $\nu_1$, and in particular at frequencies greater than $\nu_2$, the emissivity of the leaky box is significantly greater than  we would expect from electrons in a homogeneous magnetic field, i.e. from region 1 alone.

\subsection{Spectral indices}

The synchrotron spectrum arising from relativistic electrons with a power-law energy distribution within a homogeneous magnetic field can be described by a power-law with an exponential break at high frequencies where radiative losses have create a high-energy cut-off in the electron distribution. It has been pointed out \citep{kra93,rka94} that the spectra measured along the lobes of well-resolved radio galaxies show considerably more `curvature' than predicted by this simple model. The spectral curvature is best illustrated in a radio `colour-colour' plot. 

We define the spectral index between two frequencies $\nu_{\rm a}$ and $\nu_{\rm b}$ with $\nu_{\rm a} > \nu_{\rm b}$ as 
\begin{equation}
\alpha_{\rm a}^{\rm b} = \log \left( \frac{\epsilon_{\rm a}}{\epsilon _{\rm b}} \right) / \log \left( \frac{\nu_{\rm a}}{\nu_{\rm b}} \right),
\end{equation}
where $\epsilon _{\rm a}$ and $\epsilon _{\rm b}$ are the emissivities at the relevant frequencies. For observations at three frequencies, here we use 0.33, 5 and 15\,GHz, we can calculate two spectral indices and then plot these against each other in a radio colour-colour diagram. 

\begin{figure}
\centerline{
\includegraphics[width=8.45cm]{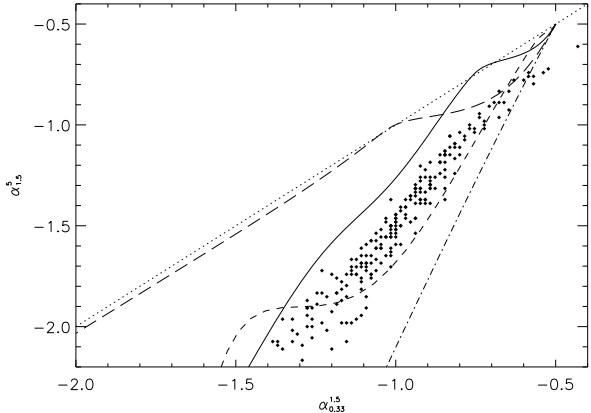}}
\caption{Colour-colour diagram for the simple leaky box with various diffusion rates $h_1$. Solid line: Evolution of the spectrum for $h_1 = 0.2 / t_0$, this is the case for which the spectra are shown in Figures \ref{sp05} to \ref{sp4}. Short-dashed line: $h_1 = 0.02 / t_0$. Long-dashed line: $h_1 = 2/ t_0$. The dotted line shows the case of a pure power-law spectrum, while the dot-dashed line applies for a homogeneous magnetic field in the emitting volume. The points are derived from observations of the hot spots and lobes of Cygnus A (\citealt{rka94}, based on the observations of \citealt{cpdl91}).}
\label{colour}
\end{figure}

In the case of a pure power-law energy distribution for the electrons without a high-energy cut-off or radiative losses we expect $\alpha_{0.33}^{1.5} = \alpha _{1.5}^{5}$. This is shown as the dotted line in the colour-colour plot in Figure \ref{colour}. The track of a population of electrons evolving in a homogeneous magnetic field is shown by the dot-dashed line. The evolution for this and any of the following tracks proceeds from the top right corner towards the left and down. The radiative losses move the high-energy cut-off to progressively lower energies and the resulting exponential drop in the spectrum causes the steep track in the colour-colour plot, since the spectral index between the higher frequencies always samples a steeper part of the exponential decay.

Taking spectra at various points along the lobes of several nearby radio galaxies, \citet{kra93} and \citet{rka94} noted that these were mainly located in the colour-colour diagram between the dotted line and the dot-dashed line. The data points in Figure \ref{colour} reproduce the results of \citep{rka94} for Cygnus A. Thus these spectra are inconsistent with spectral evolution in a homogeneous magnetic field, even in the case of inefficient pitch angle scattering \citep{kra93}. 

Figure \ref{colour} also shows the tracks for the simple leaky box model for various diffusion rates. Depending on the diffusion rate chosen, the tracks pass through the region occupied by the observational data in the colour-colour plot. The simple leaky box model is therefore consistent with observations and can reproduce the observed curvature of the radio spectra of the lobes of radio galaxies. Note however the data points indicating the regions of the lobes with the flattest spectra in the upper right-hand corner of Figure \ref{colour}. They cannot be explained with the simple leaky box model discussed here. They may require an energy distribution of the electrons flatter than $\gamma _0^{-2}$ at injection time. 

The `plateaus' seen in the evolutionary tracks through the colour-colour plot arise when one of the breaks or cut-offs in the energy distribution of the electrons passes by the lowest observing frequency. The first (upper) plateau is caused by the break at $\nu_3$. After $\nu_3$ has passed the lowest of the three observing frequency, the spectrum between them is increasingly dominated by the emission from region 2. The point where the contribution from region 2 starts exceeding that from region 1 moves from low to high frequencies. This effect therefore first leads to a steepening of the spectrum at the low observing frequencies (see Figure \ref{sp05}, and thus creates the plateau in the colour-colour plot. The first plateau occurs at flatter spectral indices for smaller diffusion rates. The plateau is also more pronounced for larger diffusion rates, so that it is difficult to see this first plateau for the lowest diffusion rate, $h_1 = 0.02 t_0$, in Figure \ref{colour}. Both effects are due to the reduced importance of the emission from region 2 compared to that from region 1 for small diffusion rates. For $h_1=0.2 t_0$ the evolutionary track even crosses the line where $\alpha_{0.33}^{1.5} = \alpha _{1.5}^{5}$ at the first plateau, so that the spectrum is steeper at lower frequencies than at higher frequencies. This spectral behaviour cannot be explained in the standard model of spectral evolution in a homogeneous magnetic field.

The second (lower) plateau is associated with the break at $\nu_1$. This break affects the emission of region 1 only and so the plateau is more pronounced for cases where the contribution of region 1 to the overall emission is most important, i.e. for low diffusion rates. As expected, the evolutionary tracks for very low diffusion rates approach the line for evolution in a homogeneous field, because the emission of region 1 dominates. For high diffusion rates the evolutionary track approximates the line $\alpha_{0.33}^{1.5} = \alpha _{1.5}^{5}$, because the emission from region 2 dominates in this case and radiative losses at the high-energy end of the electron spectrum in this region are compensated by leakage from region 1. Also, radiative losses in region 1 do not significantly affect the spectrum in the range shown in Figure \ref{colour} for this case, because the spectrum evolves rapidly along the plotted track. The position of the plateaus may be taken as rough indicators of time, since the break frequencies evolve in exactly the same way in each of the cases shown here. 

\subsection{Spectral ageing}

In the standard model of spectral evolution in a homogeneous magnetic field the position of the exponential cut-off in frequency space is used to determine the age of the emitting electron population. If the radio spectrum arises from a two-field plasma with leakage between the two regions, then the results from spectral ageing may be misleading. In fact, we will show below that the standard spectral ageing approach may produce spectra that fit observations very well, but predict spectral ages that are considerably shorter than the evolution time of the two-field plasma. 

To study this effect we use the overall spectrum of the simple leaky box with $h_1 =0.2 t_0$ and $t=t_0$ shown in Figure \ref{sp1}. We then employ standard spectral ageing techniques to derive the spectral age of this spectrum. For this analysis we follow the method discussed in \citet{al87} and \citet{pa87}. 

We assume that the spectrum produced by the simple leaky box model is observed at five radio frequencies. The flux at the two lowest frequencies, in our case 10 and 38\,MHz, are used to derive the low-frequency spectral index, $\alpha$. Since the spectrum arising from the evolution of the electron distribution in a homogeneous field predicts a power-law spectrum at low frequencies, the exponent of the underlying power-law electron energy distribution can be determined as $p=2 \alpha +1$.  In our case we find $p \sim 2.3$. Using $p$ and the `observed' emissivity at 38\,MHz we derive the strength of the magnetic field under the assumption of minimum energy conditions in the emitting plasma \citep[e.g.][]{ml94}. We find $B_{\rm min} = 22\,\mu$G which is between the magnetic field strength in the high-field region, $B_2 = 58\,\mu$G, and that in the low-field region, $B_1 = 5.8\,\mu$G, at the time of observation, $t=t_0$.

The strength of the magnetic field is usually assumed to be constant for estimates of spectral ages which simplifies the determination of the high-energy cut-off in the electron energy distribution (see equation \ref{cutoff}). This simplification also neglects the adiabatic energy losses of the electrons. With these assumptions we can now calculate spectra for different ages of the electron population and minimise their $\chi^2$-deviation from the five `observational' data points to find the best-fitting spectrum and hence the spectral age, $t_{\rm spec}$. 

\begin{figure}
\centerline{
\includegraphics[width=8.45cm]{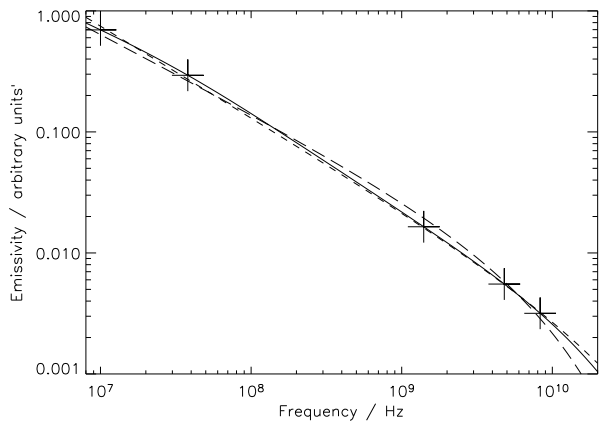}}
\caption{Standard spectral ageing techniques applied to a synthetic radio spectrum of the simple leaky model. Solid line: Spectrum of the leaky box. Long-dashed line: Spectrum from standard spectral ageing technique with $p=2.3$. Short-dashed line: Spectrum using a fixed exponent for the initial electron injection of $p=2.5$. The large crosses show the five data points used in the fitting process.}
\label{ageing}
\end{figure}

Figure \ref{ageing} shows the result of the spectral ageing analysis applied to the radio spectrum arising from the simple leaky box model as described above. Although the spectrum from the electron distribution evolving in a homogeneous magnetic field closely approximates the spectrum of the leaky box, the spectral age derived is only $3.8\times 10^6$\,years compared to the real age of $10^7$\,years. The standard spectral age determination technique underestimates the age of the relativistic plasma by a factor 2.6. Also, the exponent of the energy distribution of the injected electrons derived from the data points at low frequencies is too steep. This error can lead to a significant underestimate of the total energy content of the emitting plasma. 

In many cases no measurements of the radio flux of a source are available at low frequencies. In these cases the value of $p$ is a guess. Often $p=2.5$ is used. Figure \ref{ageing} also shows the best-fitting spectrum for this assumption. Again, the spectrum approximates the spectrum of the leaky box, but the derived spectral age is only $1.7\times 10^6$\,years, a factor 5.9 younger than the real age of the plasma. 

The reason for the underestimate of the spectral age is that the spectra of the leaky box model with $p=2$ can evolve into a single power-law implying $p>2$. In other words, the spectra of the leaky box model can mimic the spectra of electron populations with $p>2$ evolving in a homogeneous magnetic field without a high-energy cut-off. Thus the standard spectral ageing technique arrives at an exponent $p>2$ for the initial electron energy distribution and, since there is no sign of an exponential cut-off in the spectrum at high frequencies, the derived spectral ages are small. From Figure \ref{colour} it is clear that this effect is particularly strong for high diffusion rates. 

A secondary effect is the move of the spectral cut-off to higher frequencies by the stronger magnetic field in region 2 as discussed in Section \ref{spec} in connection with Figure \ref{sp4}. This effect also leads to an underestimate of the age of the electron population by the standard spectral ageing technique. 

\section{Summary}
\label{conc}

We have developed an analytical model for the evolution of the electron energy distribution in a relativistic, magnetised plasma divided into two regions with different magnetic field strengths. This leaky box model is designed to mimic the evolution of a realistic magnetised plasma in which the magnetic field concentrates in magnetic flux ropes rather than being distributed homogeneously throughout the plasma. The relativistic electrons are allowed to leak from the region with a low magnetic field strength into the high-field region, but not back again. Also, we assume that the diffusion rate associated with this leakage is independent of the electron energy. For simplicity we assume that the diffusion rate is constant and that the initially injected electron population has an energy distribution proportional to $\gamma ^{-2}$. In both regions the electrons are subject to energy losses due to adiabatic expansion, synchrotron radiation and inverse Compton scattering of the CMB. 

The evolution of the energy distribution of the electrons in the low-field region is identical to that of electrons in the commonly assumed homogeneous magnetic field, except for the integrated loss of electrons to the high-field region. As expected, the distribution develops a high-energy cut-off at $\gamma_{\rm 1, max}$ in terms of the Lorentz factor. Because of the continued leaking of electrons to the high-field region and the faster energy losses there, $\gamma_{\rm 1,max}$ is also the high-energy cut-off in the high-energy region. However, the electron energy distribution in the high-field region also develops a break at $\gamma_{\rm 2, max} < \gamma _{\rm1, max}$. The break occurs at the maximum Lorentz factor of those electrons that leaked into the high-field region at the earliest possible time which are therefore those electrons that have suffered the largest energy loss in the high-field region. 

The emission spectra of the leaky box model contain three spectral breaks associated with $\gamma _{\rm 1,max}$ and $\gamma_{\rm 2,max}$. The spectral slopes in between the breaks depend strongly on the diffusion rate between the two regions and the time of observation. In general, the resulting radio spectra are more curved than those arising from electron populations evolving in homogeneous magnetic fields. Therefore the spectra from the leaky box are consistent with observations of curved spectra in radio galaxies \citep{kra93,rka94}. More complex behaviour of the spectra is possible for the more complicated choice of a time-variable diffusion rate and/or a different energy distribution for the initially injected electrons.

We also applied standard spectral ageing techniques to the radio spectra from the leaky box model. We found that, while the standard spectral ageing model produces good fits to the spectra,  it significantly underestimates the age of the electron population. The standard spectral ageing model derives the exponent of the assumed power-law energy distribution of the radiating electrons from the spectral slope at low frequencies. The leaky box spectra show significant steepening of the spectra even at these low frequencies and thus the standard models overestimate the value of the exponent. The discrepancies between the estimated spectral ages and the estimated exponent of the power-law energy distribution and their real values is greatest for large diffusion rates between the two regions. 

Finally, the leaky box model can considerably extend the time over which a given population of relativistic electrons can be detected at a given frequency. The energy of the electrons is preserved in the low-field region. When they leak into the high-field region they light up and lose their energy quickly. However, more electrons of a given energy are available at later times to leak into the high-field region than for the case of a homogeneous magnetic field. Thus synchrotron emitting structures remain detectable for longer than usually assumed. This may help to reconcile the long dynamic ages of buoyant radio plasma in galaxy clusters with the short radiative ages derived using the standard spectral ageing model \citep[e.g.][]{cbkbf00,bm01,bkc01}.

\section*{Acknowledgments}

I would like to thank P. Alexander for very helpful discussions. Many thanks also to L. Rudnick for providing me with the observational colour-colour data of Cygnus A in electronic form and pointing out the online version of P. Tribble's paper on the same topic. I thank the referee, P. Wiita, for constructive comments that improved the paper.

\def\newblock{\hskip .11em plus .33em minus .07em}

\bibliography{crk}
\bibliographystyle{mn2e}


\end{document}